\RequirePackage{lineno}
\documentclass[aps,prl,twocolumn,showpacs,superscriptaddress,letterpaper,amsmath,amssymb]{revtex4}  

\usepackage{graphicx}
\usepackage{dcolumn}
\usepackage{bm}

\RequirePackage{xspace}
\usepackage{relsize}

\def\Kbar  {\kern 0.2em\overline{\kern -0.2em K}{}\xspace}

\def\Bbar    {\kern 0.18em\overline{\kern -0.18em B}{}\xspace}

\def\Qbar    {\kern 0.08em\overline{\kern -0.08em Q}{}\xspace}

\newcommand{\mev}{\ensuremath{\mathrm{\,Me\kern -0.1em V}}\xspace}
\newcommand{\mevc}{\ensuremath{{\mathrm{\,Me\kern -0.1em V\!/}c}}\xspace}
\newcommand{\mevcc}{\ensuremath{{\mathrm{\,Me\kern -0.1em V\!/}c^2}}\xspace}
\newcommand{\gev}{\ensuremath{\mathrm{\,Ge\kern -0.1em V}}\xspace}
\newcommand{\gevc}{\ensuremath{{\mathrm{\,Ge\kern -0.1em V\!/}c}}\xspace}
\newcommand{\gevcnospace}{\ensuremath{{\mathrm{\,Ge\kern -0.1em V\!/}c}}}
\newcommand{\gevcc}{\ensuremath{{\mathrm{\,Ge\kern -0.1em V\!/}c^2}}\xspace}

\def\missET {\mbox{\ensuremath{\not\!\! E_T}}}

\begin{document}
\bibliographystyle{apsrev}

\title{{Search for new physics in high $p_T$ like-sign dilepton events at CDF~II}}

\affiliation{Institute of Physics, Academia Sinica, Taipei, Taiwan 11529, Republic of China} 
\affiliation{Argonne National Laboratory, Argonne, Illinois 60439, USA} 
\affiliation{University of Athens, 157 71 Athens, Greece} 
\affiliation{Institut de Fisica d'Altes Energies, ICREA, Universitat Autonoma de Barcelona, E-08193, Bellaterra (Barcelona), Spain} 
\affiliation{Baylor University, Waco, Texas 76798, USA} 
\affiliation{Istituto Nazionale di Fisica Nucleare Bologna, $^{aa}$University of Bologna, I-40127 Bologna, Italy} 
\affiliation{University of California, Davis, Davis, California 95616, USA} 
\affiliation{University of California, Irvine, California 92627, USA}
\affiliation{University of California, Los Angeles, Los Angeles, California 90024, USA} 
\affiliation{Instituto de Fisica de Cantabria, CSIC-University of Cantabria, 39005 Santander, Spain} 
\affiliation{Carnegie Mellon University, Pittsburgh, Pennsylvania 15213, USA} 
\affiliation{Enrico Fermi Institute, University of Chicago, Chicago, Illinois 60637, USA}
\affiliation{Comenius University, 842 48 Bratislava, Slovakia; Institute of Experimental Physics, 040 01 Kosice, Slovakia} 
\affiliation{Joint Institute for Nuclear Research, RU-141980 Dubna, Russia} 
\affiliation{Duke University, Durham, North Carolina 27708, USA} 
\affiliation{Fermi National Accelerator Laboratory, Batavia, Illinois 60510, USA} 
\affiliation{University of Florida, Gainesville, Florida 32611, USA} 
\affiliation{Laboratori Nazionali di Frascati, Istituto Nazionale di Fisica Nucleare, I-00044 Frascati, Italy} 
\affiliation{University of Geneva, CH-1211 Geneva 4, Switzerland} 
\affiliation{Glasgow University, Glasgow G12 8QQ, United Kingdom} 
\affiliation{Harvard University, Cambridge, Massachusetts 02138, USA} 
\affiliation{Division of High Energy Physics, Department of Physics, University of Helsinki and Helsinki Institute of Physics, FIN-00014, Helsinki, Finland} 
\affiliation{University of Illinois, Urbana, Illinois 61801, USA} 
\affiliation{The Johns Hopkins University, Baltimore, Maryland 21218, USA} 
\affiliation{Institut f\"{u}r Experimentelle Kernphysik, Karlsruhe Institute of Technology, D-76131 Karlsruhe, Germany} 
\affiliation{Center for High Energy Physics: Kyungpook National University, Daegu 702-701, Korea; Seoul National University, Seoul 151-742, Korea; Sungkyunkwan University, Suwon 440-746, Korea; Korea Institute of Science and Technology Information, Daejeon 305-806, Korea; Chonnam National University, Gwangju 500-757, Korea; Chonbuk National University, Jeonju 561-756, Korea} 
\affiliation{Ernest Orlando Lawrence Berkeley National Laboratory, Berkeley, California 94720, USA} 
\affiliation{University of Liverpool, Liverpool L69 7ZE, United Kingdom} 
\affiliation{University College London, London WC1E 6BT, United Kingdom} 
\affiliation{Centro de Investigaciones Energeticas Medioambientales y Tecnologicas, E-28040 Madrid, Spain} 
\affiliation{Massachusetts Institute of Technology, Cambridge, Massachusetts 02139, USA} 
\affiliation{Institute of Particle Physics: McGill University, Montr\'{e}al, Qu\'{e}bec, Canada H3A~2T8; Simon Fraser University, Burnaby, British Columbia, Canada V5A~1S6; University of Toronto, Toronto, Ontario, Canada M5S~1A7; and TRIUMF, Vancouver, British Columbia, Canada V6T~2A3} 
\affiliation{University of Michigan, Ann Arbor, Michigan 48109, USA} 
\affiliation{Michigan State University, East Lansing, Michigan 48824, USA}
\affiliation{Institution for Theoretical and Experimental Physics, ITEP, Moscow 117259, Russia}
\affiliation{University of New Mexico, Albuquerque, New Mexico 87131, USA} 
\affiliation{Northwestern University, Evanston, Illinois 60208, USA} 
\affiliation{The Ohio State University, Columbus, Ohio 43210, USA} 
\affiliation{Okayama University, Okayama 700-8530, Japan} 
\affiliation{Osaka City University, Osaka 588, Japan} 
\affiliation{University of Oxford, Oxford OX1 3RH, United Kingdom} 
\affiliation{Istituto Nazionale di Fisica Nucleare, Sezione di Padova-Trento, $^{bb}$University of Padova, I-35131 Padova, Italy} 
\affiliation{LPNHE, Universite Pierre et Marie Curie/IN2P3-CNRS, UMR7585, Paris, F-75252 France} 
\affiliation{University of Pennsylvania, Philadelphia, Pennsylvania 19104, USA}
\affiliation{Istituto Nazionale di Fisica Nucleare Pisa, $^{cc}$University of Pisa, $^{dd}$University of Siena and $^{ee}$Scuola Normale Superiore, I-56127 Pisa, Italy} 
\affiliation{University of Pittsburgh, Pittsburgh, Pennsylvania 15260, USA} 
\affiliation{Purdue University, West Lafayette, Indiana 47907, USA} 
\affiliation{University of Rochester, Rochester, New York 14627, USA} 
\affiliation{The Rockefeller University, New York, New York 10065, USA} 
\affiliation{Istituto Nazionale di Fisica Nucleare, Sezione di Roma 1, $^{ff}$Sapienza Universit\`{a} di Roma, I-00185 Roma, Italy} 

\affiliation{Rutgers University, Piscataway, New Jersey 08855, USA} 
\affiliation{Texas A\&M University, College Station, Texas 77843, USA} 
\affiliation{Istituto Nazionale di Fisica Nucleare Trieste/Udine, I-34100 Trieste, $^{gg}$University of Udine, I-33100 Udine, Italy} 
\affiliation{University of Tsukuba, Tsukuba, Ibaraki 305, Japan} 
\affiliation{Tufts University, Medford, Massachusetts 02155, USA} 
\affiliation{University of Virginia, Charlottesville, Virginia 22906, USA}
\affiliation{Waseda University, Tokyo 169, Japan} 
\affiliation{Wayne State University, Detroit, Michigan 48201, USA} 
\affiliation{University of Wisconsin, Madison, Wisconsin 53706, USA} 
\affiliation{Yale University, New Haven, Connecticut 06520, USA} 
\author{T.~Aaltonen}
\affiliation{Division of High Energy Physics, Department of Physics, University of Helsinki and Helsinki Institute of Physics, FIN-00014, Helsinki, Finland}
\author{B.~\'{A}lvarez~Gonz\'{a}lez$^w$}
\affiliation{Instituto de Fisica de Cantabria, CSIC-University of Cantabria, 39005 Santander, Spain}
\author{S.~Amerio}
\affiliation{Istituto Nazionale di Fisica Nucleare, Sezione di Padova-Trento, $^{bb}$University of Padova, I-35131 Padova, Italy} 

\author{D.~Amidei}
\affiliation{University of Michigan, Ann Arbor, Michigan 48109, USA}
\author{A.~Anastassov}
\affiliation{Northwestern University, Evanston, Illinois 60208, USA}
\author{A.~Annovi}
\affiliation{Laboratori Nazionali di Frascati, Istituto Nazionale di Fisica Nucleare, I-00044 Frascati, Italy}
\author{J.~Antos}
\affiliation{Comenius University, 842 48 Bratislava, Slovakia; Institute of Experimental Physics, 040 01 Kosice, Slovakia}
\author{G.~Apollinari}
\affiliation{Fermi National Accelerator Laboratory, Batavia, Illinois 60510, USA}
\author{J.A.~Appel}
\affiliation{Fermi National Accelerator Laboratory, Batavia, Illinois 60510, USA}
\author{A.~Apresyan}
\affiliation{Purdue University, West Lafayette, Indiana 47907, USA}
% ...
\collaboration{CDF Collaboration\footnote{With visitors from $^a$Istituto Nazionale di Fisica Nucleare, Sezione di Cagliari, 09042 Monserrato (Cagliari), Italy,
$^b$University of CA Irvine, Irvine, CA  92697, USA,
$^c$University of CA Santa Barbara, Santa Barbara, CA 93106, USA,
$^d$University of CA Santa Cruz, Santa Cruz, CA  95064, USA,
$^e$CERN,CH-1211 Geneva, Switzerland,
$^f$Cornell University, Ithaca, NY  14853, USA, 
$^g$University of Cyprus, Nicosia CY-1678, Cyprus, 
$^h$Office of Science, U.S. Department of Energy, Washington, DC 20585, USA,
$^i$University College Dublin, Dublin 4, Ireland,
$^j$University of Fukui, Fukui City, Fukui Prefecture, Japan 910-0017,
$^k$Universidad Iberoamericana, Mexico D.F., Mexico,
$^l$Iowa State University, Ames, IA  50011, USA,
$^m$University of Iowa, Iowa City, IA  52242, USA,
$^n$Kinki University, Higashi-Osaka City, Japan 577-8502,
$^o$Kansas State University, Manhattan, KS 66506, USA,
$^p$University of Manchester, Manchester M13 9PL, United Kingdom,
$^q$Queen Mary, University of London, London, E1 4NS, United Kingdom,
$^r$University of Melbourne, Victoria 3010, Australia,
$^s$Muons, Inc., Batavia, IL 60510, USA,
$^t$Nagasaki Institute of Applied Science, Nagasaki, Japan, 
$^u$National Research Nuclear University, Moscow, Russia,
$^v$University of Notre Dame, Notre Dame, IN 46556, USA,
$^w$Universidad de Oviedo, E-33007 Oviedo, Spain, 
$^x$Texas Tech University, Lubbock, TX  79609, USA,
$^y$Universidad Tecnica Federico Santa Maria, 110v Valparaiso, Chile,
$^z$Yarmouk University, Irbid 211-63, Jordan,
$^{hh}$On leave from J.~Stefan Institute, Ljubljana, Slovenia, 
}}
\noaffiliation

\begin{abstract}
We present a search for new physics in events with two high $p_T$
 leptons of the same electric charge, using data with an integrated
 luminosity of 6.1 fb$^{-1}$. The observed data are consistent with
 standard model predictions. We set 95\% C.L. lower limits on the mass of
 doubly-charged scalars decaying to like-sign dileptons, $m_{H^{\pm\pm}}>
 190-245$ GeV/$c^2$, depending on the decay mode and coupling.
\end{abstract}



\maketitle

\date{\today}

A wide variety of models of new physics predict events with two
like-sign leptons, a signature which has very low backgrounds from the
standard model.    Examples include doubly-charged Higgs
bosons~\cite{hpptheory}, supersymmetry~\cite{susy}, heavy
neutrinos~\cite{heavynu}, like-sign top quark production~\cite{sstt}, and fourth-generation quarks~\cite{reviews}.

CDF examined the like-sign dilepton data with integrated luminosity of
110 pb$^{-1}$ in Run I~\cite{matt_jane_run1}
and 1 fb$^{-1}$ in Run II~\cite{lsdil_previous}, observing in Run II a modest
excess of events above the standard model expectation (44 observed, $33.2 \pm
4.7$ expected).

In this Letter, we present a study of events with like-sign
dileptons with an integrated luminosity of 6.1 fb$^{-1}$ collected by
the CDF~II detector.  We search
for a localized excess of events in a model-independent manner by
comparing the observed events to the standard model prediction using a 
Kolmogorov-Smirnov test in several kinematic variables and assessing
the statistical consistency.  In addition, we set limits on a specific
model: pair production of doubly-charged
scalars which decay to two like-sign charged
leptons~\cite{hpptheory}. These limits supersede those from CDF in 240 pb$^{-1}$~\cite{cdfhpp} and are
 stronger than those from D0 in 1.1 fb$^{-1}$~\cite{d0hpp} and CMS in
 36 pb$^{-1}$~\cite{cmshpp} by an order of magnitude. A companion
article~\cite{lsdilprd} includes interpretations for like-sign top quark production
and supersymmetric processes.

Events were recorded by CDF II~\cite{cdf,coordinates}, a general purpose detector
designed to study collisions at the Fermilab Tevatron
$p\overline{p}$ collider at $\sqrt{s}=1.96$ TeV. A charged-particle tracking system immersed
in a 1.4~T magnetic field consists of a silicon microstrip tracker and
a drift chamber. Electromagnetic and hadronic calorimeters surround
the tracking system and measure particle energies. Drift chambers
located outside the calorimeters detect muons. We examine data taken
between August 2002 and September 2010, corresponding to an integrated luminosity of
6.1 fb$^{-1}$. 

The data acquisition system is triggered by $e$ or $\mu$ candidates~\cite{leptonid}
with transverse momentum\cite{coordinates}, $p_T$,  greater than $
18$ GeV$/c$.  Electrons and muons are reconstructed offline and selected if they
have a pseudorapidity\cite{coordinates}, $\eta$, magnitude less than
1.1, $p_T \ge 20$ GeV$/c$ and satisfy the standard CDF identification and
isolation requirements~\cite{leptonid}.  An additional requirement is
made to suppress electrons from photon conversions, by rejecting
electron candidates with a collinear intersecting reconstructed track. Jets are reconstructed in the
calorimeter using the {\sc jetclu}~\cite{jetclu}  algorithm with a
clustering radius of 0.4 in azimuth-pseudorapidity space~\cite{coordinates} and calibrated~\cite{jes}. Jets are
selected if they have $p_T \ge 15$ GeV$/c$ and $|\eta| < 2.4$.
Missing transverse momentum~\cite{metdef}, $\missET$,
is reconstructed using fully corrected
calorimeter and muon information~\cite{leptonid}.

We select events with at least two isolated leptons (electrons or
muons), two of which have the same electric charge.  The leading lepton must have $p_T > 20$
GeV$/c$, $|\eta|<1.1$ and be isolated in both the calorimeter and the
tracker. The second lepton must satisfy the same requirements, with
the exception that it needs only have $p_T > 10$ GeV$/c$.  We require that the two leptons
come from the same primary vertex and have a dilepton invariant mass
$m_{\ell\ell}$ of at least 25 GeV$/c^{2}$ to reduce backgrounds from
pair production of bottom quarks. Finally, we reject events
with three or more leptons if they contain a pair of opposite-sign
leptons or like-signed electrons in the window, $m_{\ell\ell} \in
[86,96]$ GeV$/c^2$. Like-signed electrons pairs may be produced by the
radiation of a hard photon, see below, which is negligible for muons. In each event, we calculate $H_T$, the scalar sum of the lepton
$p_T$, the jet $E_T$ and the missing transverse momentum.

Irreducible backgrounds to the like-sign dilepton signature with prompt
like-sign leptons are rare in the SM; they are largely from $WZ$ and
$ZZ$ production.  These backgrounds are modeled using simulated events
generated by {\sc pythia}~\cite{pythia} with the detector response
simulated with a {\sc geant}-based algorithm {\sc
  cdfsim}~\cite{cdfsim}.

The dominant reducible background comes from $W+$jets production or $t\bar{t}$
production with semi-leptonic decays, with one prompt lepton and a second lepton
 due to the semi-leptonic decay of a $b$- or $c$-quark meson.  This (``fake'') background is described using a
lepton misidentification model from inclusive jet data
applied to $W+$jet events, validated in orthogonal jet samples and in
events with like-sign dileptons but low invariant mass: $m_{\ell\ell}
\in [15,25]$ GeV/$c^2$.

The second largest source of background comes from processes which
produce electron-positron pairs; either the electron or positron emits
a hard photon leading to an asymmetric conversion
(e.g. $e^-_{hard}\rightarrow e^-_{soft}\gamma\rightarrow e^-_{soft}
e^-_{soft}  e^+_{hard}$) where the track for the $e^+_{hard}$ determines
the charge.   This mechanism is well-described by the detector simulation,
and is validated in events with like-sign electron pairs which have a
conversion-tagged electron.  The major contributions via this mechanism are from
$Z/\gamma^*$+jets and $t\bar{t}$ production with fully leptonic
decays. Estimates of the backgrounds from $Z/\gamma^*$+jets processes
are modeled using simulated events generated by {\sc pythia} normalized to data in opposite-sign events.
The detector response for both $Z$+jets and $t\overline{t}$ processes
is evaluated using {\sc cdfsim}, where, to avoid double-counting, the
like-sign leptons are required to originate from the $W$ or $Z$ boson
decays rather than from misidentified jets.  

An additional contribution to the background is due to associated production of a $W$
boson with a prompt photon. If the $W$ boson decays to an electron (muon) and
the photon converts too early to be identified as a conversion, the
event can be reconstructed with a like-sign $ee$
($e\mu$) signature.  The rate of $W\gamma$ production the efficiency
for finding conversions is validated in a sample of like-sign dilepton events with a
conversion-tagged electron.

Backgrounds from charge-mismeasurement are insignificant, as the
charge of a particle with momentum of 100 GeV/$c$ is typically
determined with a significance greater than 5$\sigma$~\cite{wandz}.

The dominant systematic uncertainty is the 50\% uncertainty of the
lepton misidentification rate, due to possible
contamination of leptons from $W$ and $Z$ boson decays in the inclusive jet
data. This gives a 20\% uncertainty on the total background. Additional uncertainties are due to
the jet energy scale~\cite{jes}, contributions from additional interactions, and descriptions of
initial and final state radiation~\cite{rad} and uncertainties in the
parton distribution functions~\cite{pdfs1,pdfs2}.

\begin{figure}
\centering
\includegraphics[width=0.49\linewidth]{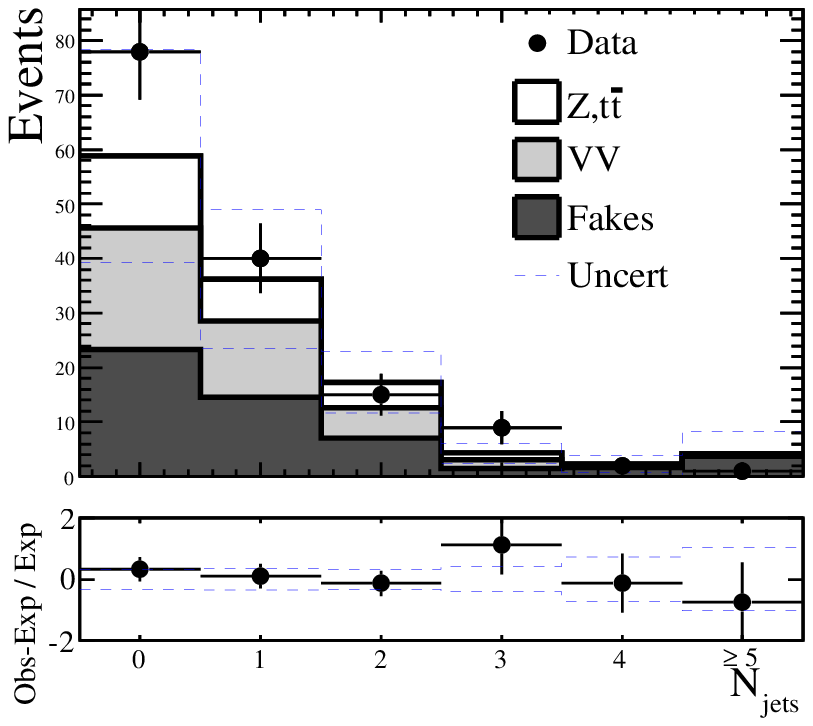}
\includegraphics[width=0.49\linewidth]{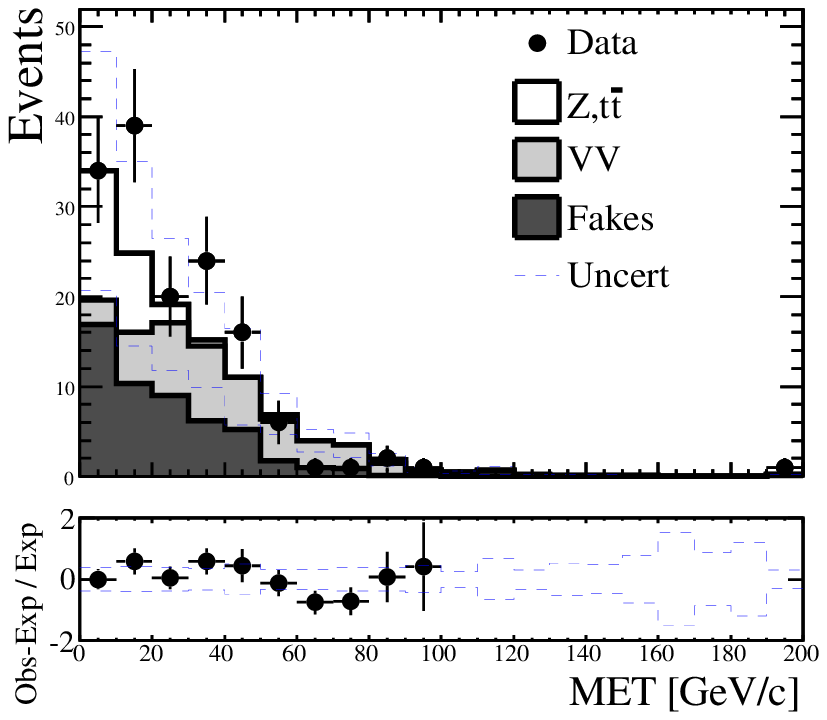}\\
\includegraphics[width=0.49\linewidth]{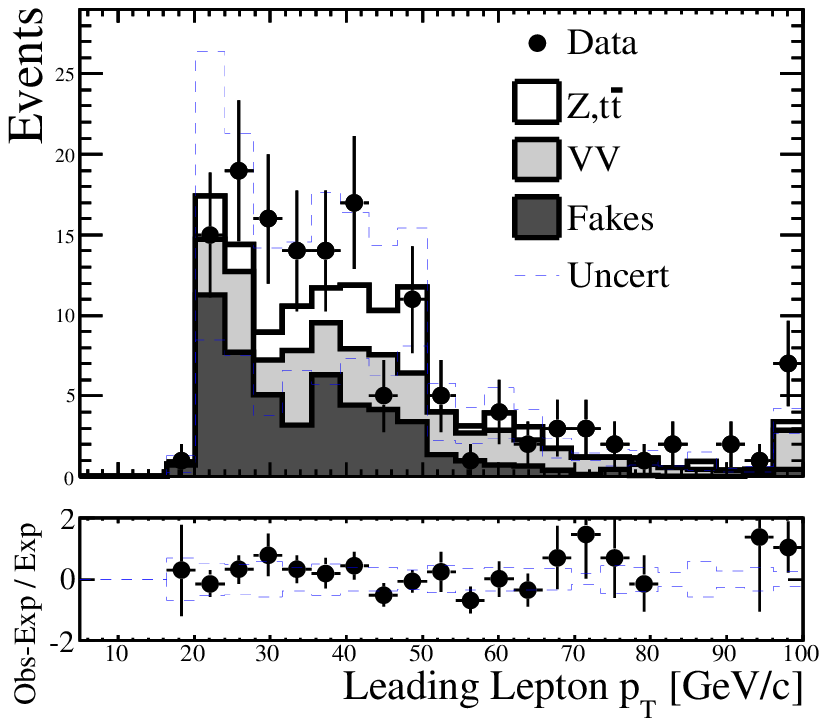}
\includegraphics[width=0.49\linewidth]{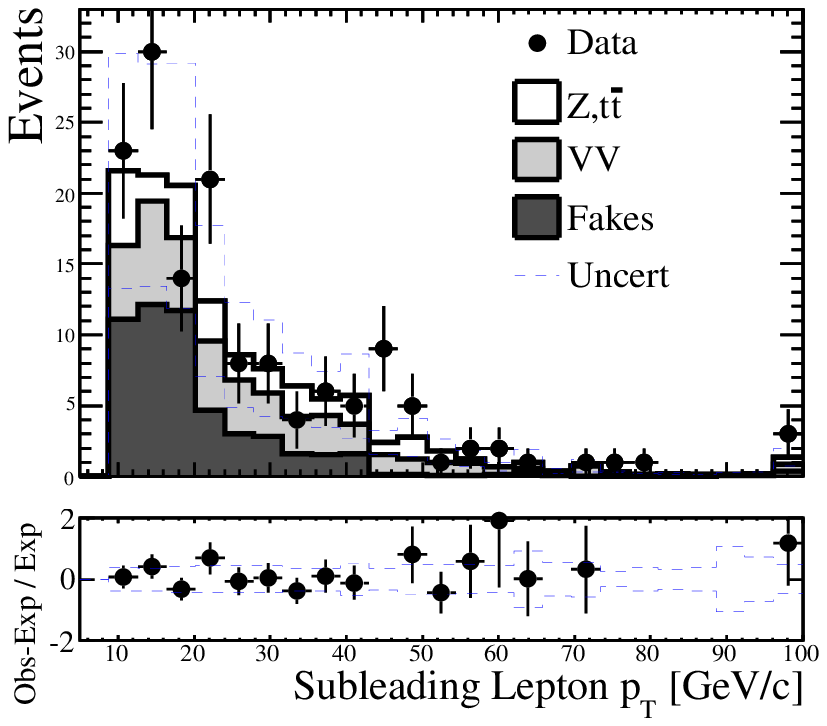}
\caption{ Distribution of jet multiplicity, missing transverse momentum,
  leading lepton $p_T$ and sub-leading lepton $p_T$ in observed like-sign
  dilepton events and expected backgrounds. The $VV$ contribution
  includes $WW,WZ,ZZ$ and $W\gamma$. }
\label{fig:sskin}
\end{figure}

Table~\ref{tab:yields} shows the observed and predicted event
yields. Figure~\ref{fig:sskin} shows kinematic distributions of observed and predicted like-sign lepton events.

\begin{table}
\caption{Predicted and observed event yields in like-sign lepton
  events. Uncertainties included statistical and systematic
  contributions. Entries written as --- are negigible.}
\begin{center}
\begin{tabular}{lrrrr}
\hline\hline
Process& Total $\ell\ell$& $\mu\mu$& ee& $e\mu$\\\hline
$t\bar{t}$ & $0.1 \pm 0.1$ &---& --- & $0.1 \pm 0.1$\\
$Z\rightarrow \ell\ell$ & $26.6 \pm 3.4$ & --- & $17.0 \pm 2.8$ & $9.7 \pm 2.1$\\
$WW,WZ,ZZ$ & $28.4 \pm 1.4$ & $7.9 \pm 0.9$ & $6.0 \pm 0.4$ & $14.5 \pm 0.8$\\
$W(\rightarrow \ell\nu)\gamma$ & $16.2 \pm 2.4$ & --- & $8.1 \pm 1.8$ & $8.0 \pm 1.8$\\
Fake Leptons & $51.6 \pm 24.2$ & $8.2 \pm 5.3$ & $22.1 \pm 8.9$ & $21.3 \pm 10.6$\\
\hline
 Total & $123.0 \pm 24.6$& $16.1 \pm 5.4$& $53.3 \pm 9.5$& $53.6 \pm 10.9$\\\hline
 Data & 145& 14& 66& 65\\\hline\hline
\end{tabular}
\end{center}
\label{tab:yields}
\end{table}

We calculate the maximum Kolmogorov-Smirnov (KS) distance for each the
distributions $m_{\ell\ell}$, $\missET$, $N_{jets}$, lepton $p_T$ and
$H_T$. A large KS distance value would indicate a localized excess in one of these
variables, though this test is not sensitive to discrepancies in the total yield. In each case,
the standard model $p$-value (probability to observe a result at least
this discrepant from the standard model) does not indicate significant deviation from the
background-only hypothesis; see Table~\ref{tab:KSresults}.

This larger dataset does not show evidence of the excess seen in the
previous analysis~\cite{lsdil_previous} that was based on 1~fb$^{-1}$
of integrated luminosity. The  background from misidentified leptons was
calculated using a different technique, which gives a larger
estimate in the original dataset than the previous analysis, though
consistent within systematic uncertainties.

\begin{table}
\caption{Results of KS-distance test for standard model prediction.
The maximum KS distance and corresponding $p$-value is given for several
kinematic distributions presented in this analysis.}
\begin{center}
\begin{tabular}{lcccc}
\hline\hline
Distribution                    & Total $\ell\ell$  & $ee$ &$\mu\mu$& $e\mu$\\ \hline
$m_{\ell\ell}$ & 0.11 (79\%) & 0.22 (47\%) & 0.23 (46\%) & 0.30 (59\%)\\ 
$\missET$ & 0.19 (34\%) & 0.23 (27\%) & 0.24 (32\%) & 0.21 (69\%)\\ 
$N_{jets}$ & 0.19 (56\%) & 0.31 (31\%) & 0.20 (57\%) & 0.21 (84\%)\\ 
Lepton 1 $p_T$ & 0.16 (49\%) & 0.18 (47\%) & 0.25 (30\%) & 0.26 (60\%)\\ 
Lepton 2 $p_T$ & 0.12 (66\%) & 0.21 (41\%) & 0.23 (33\%) & 0.40 (33\%)\\ 
$H_T$ & 0.15 (45\%) & 0.22 (34\%) & 0.22 (32\%) & 0.24 (58\%)\\ 
\hline\hline
\end{tabular}
\end{center}
\label{tab:KSresults}
\end{table}

\begin{figure}
\centering
\includegraphics[width=0.7\linewidth]{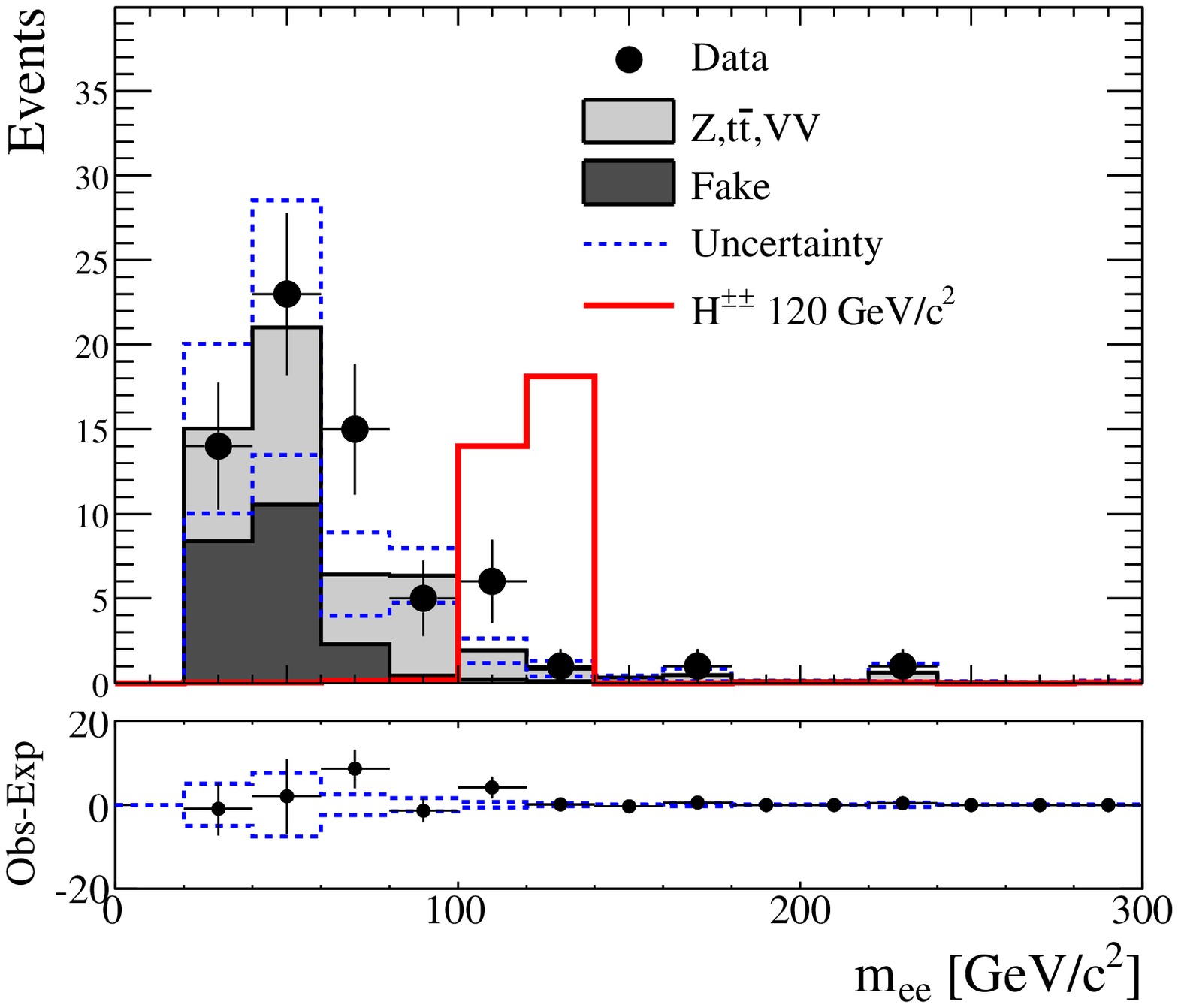}
\includegraphics[width=0.7\linewidth]{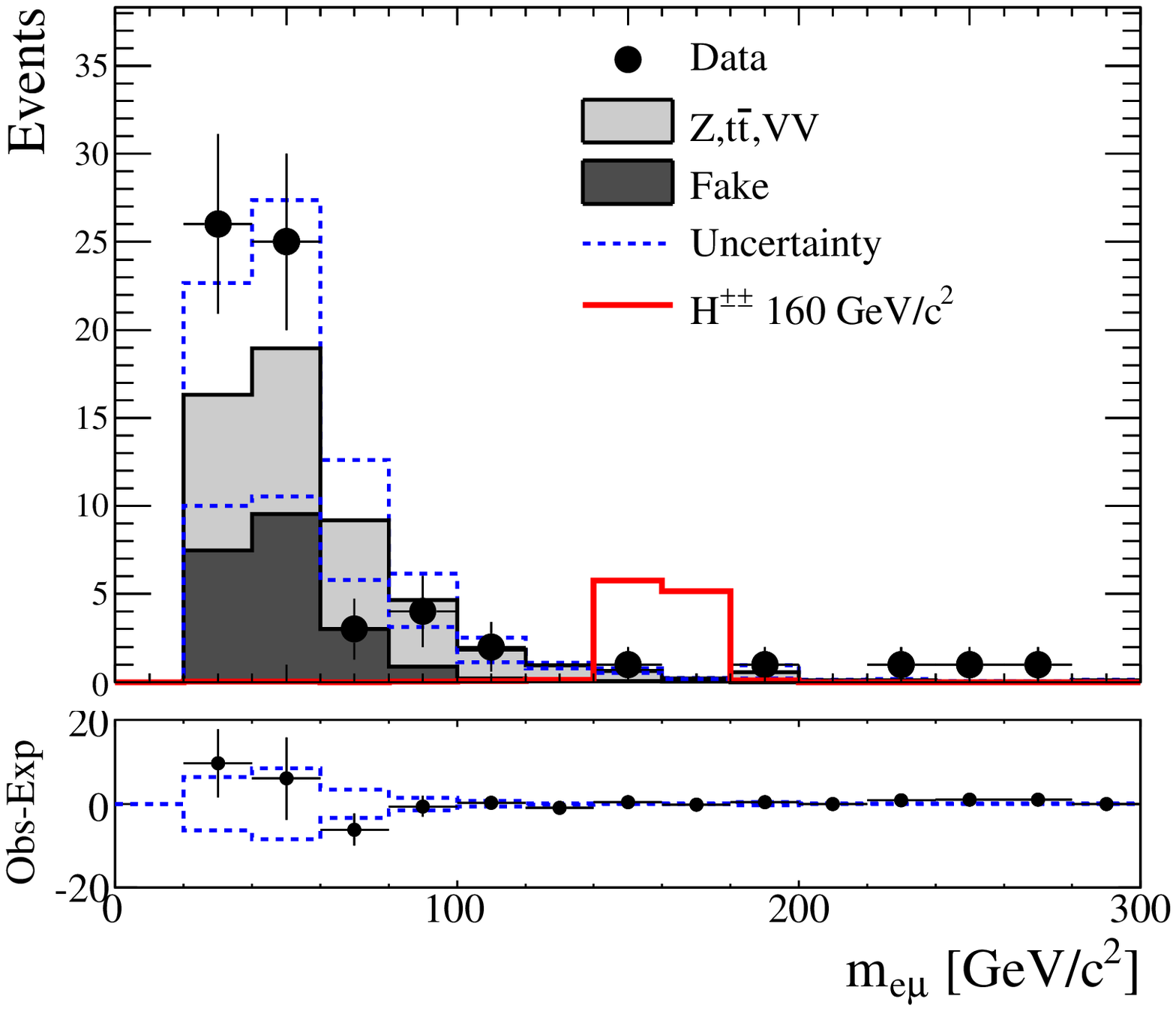}
\includegraphics[width=0.7\linewidth]{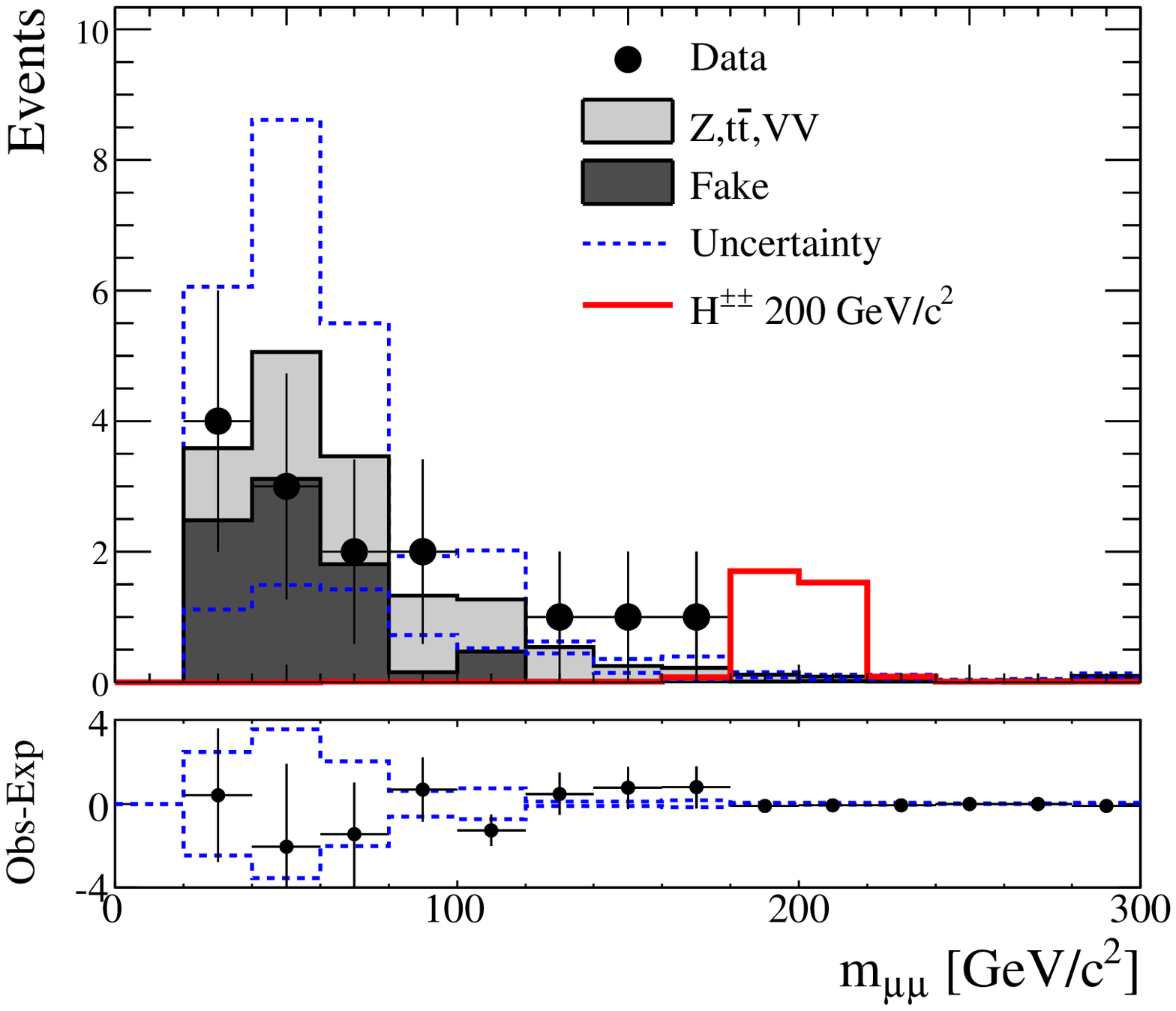}
\caption{ The observed  and expected standard model spectra in the
$ee,\mu\mu$ and $e\mu$ channels. The doubly-charged Higgs boson signal
is shown for typical masses.  The $VV$ contribution
  includes $WW,WZ,ZZ$ and $W\gamma$.}
\label{fig:mll}
\end{figure}

Observing no excess, we report our sensitivity in terms of limits on doubly-charge scalar bosons decaying to like-sign electron pairs, muon pairs or
electron-muon pairs.  Simulated events are generated with {\sc
  madevent}~\cite{madgraph}, showering and hadronization is performed by {\sc pythia}
passed through the CDF~II full detector simulation. Figure~\ref{fig:mll}
shows the observed and expected standard model spectra in the
$ee,\mu\mu$ and $e\mu$ channels.

 The largest uncertainties on the signal model are due to
energy resolution and lepton identification efficiencies, which are
minor compared to the background uncertainties. In each case, we treat
the unknown underlying quantity as a nuisance parameter and measure the distortion of the
dilepton mass spectrum for positive and negative fluctuations. 

\begin{figure}
\centering
\includegraphics[width=0.8\linewidth]{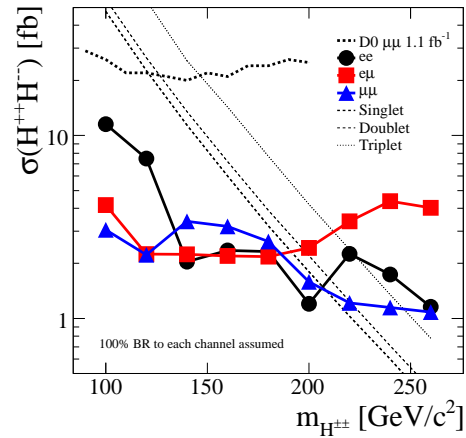}
\caption{ Observed upper limits at 95\% C.L. on the production cross-section for
  doubly-charged Higgs, assuming 100\% branching fraction to
  $ee,\mu\mu$ or $e\mu$, compared to results from D0~\cite{d0hpp}. Also shown are next-to-leading-order
  theoretical calculations of the cross-section, assuming the Higgs is
  a member of a singlet, doublet or triplet.}
\label{fig:hpp}
\end{figure}

The dilepton mass spectrum is in good agreement with the standard
model prediction, and we calculate 95\% confidence level upper limits on the production cross section of doubly-charged
Higgs bosons, using
frequentist statistics with the unified ordering
scheme~\cite{feldcous}.  The $Z/\gamma^*$ coupling and therefore production cross-section of the
doubly-charged Higgs boson depends on whether it is a member of a
singlet, doublet or triplet, as shown in Fig.~\ref{fig:hpp} and
Tables~\ref{tab:hpp} and~\ref{tab:hm}.

\begin{table}
\begin{center}
\caption{For various Higgs masses, the NLO cross sections for singlet
  ($\sigma_1$), doublet ($\sigma_2$), triplet ($\sigma_3$) production,
  expected and observed 95\% C.L. limits in the $ee,e\mu$ and $\mu\mu$ channels. All
  cross-sections are in femtobarns.}
\begin{tabular}{llllrrrrrr}
\hline\hline
$m_{H^{\pm\pm}}$ & \multicolumn{3}{c}{Theory} & \multicolumn{3}{c}{Observed}  & \multicolumn{3}{c}{Expected} \\
(GeV/$c^2$) & $\sigma_1$   & $\sigma_2$  &
$\sigma_3$\ \ \ \ \  & $\sigma^{95}_{ee}$   & $\sigma^{95}_{e\mu}$ &
$\sigma^{95}_{\mu\mu}$   & \ \ \ \ \ $\sigma^{95}_{ee}$   & $\sigma^{95}_{e\mu}$ & $\sigma^{95}_{\mu\mu}$  \\
\hline
100     & 48  & 55  & 120   & 12 & 4.2  & 3.1 & 5.7  & 3.8  & 2.8\\
120     & 23  & 27  & 55  & 7.4  & 2.3  & 2.2 & 3.3  & 2.6  & 2.2\\
140     & 11  & 14  & 26  & 2.0  & 2.2  & 3.4 & 2.4  & 2.2  & 1.6\\
160     & 6.0     & 7.2   & 14  & 2.4  & 2.2  & 3.2 & 1.5  & 1.9  & 1.4\\

180     & 3.2   & 3.9   & 7.7   & 2.3  & 2.2  & 2.6 & 1.5  & 1.7  & 1.5\\

200     & 1.8   & 2.2   & 4.2   & 1.2  & 2.4  & 1.6 & 1.5  & 1.4  & 1.5\\

220     & 1.0  & 1.2  & 2.4  & 2.3  & 3.4  & 1.2 & 1.4  & 1.1  & 1.4\\

240     & 0.6  & 0.7  & 1.4  & 1.7  & 4.4  & 1.2 & 1.4  & 1.1  & 1.3\\

260     & 0.3  & 0.4  & 0.8  & 1.2  & 4.0  & 1.1 & 1.5  & 1.0  & 1.2\\
\hline\hline
\end{tabular}
\label{tab:hpp}
\end{center}
\end{table}

\begin{table}
\caption{Lower limits at 95\% C.L. on $H^{\pm\pm}$ masses by channel,for singlet,
  doublet and triplet theories. All in units of GeV/$c^{2}$.}
\begin{tabular}{lrrr}
\hline\hline
& \multicolumn{3}{c}{Theory} \\
Channel & Triplet & Doublet & Singlet \\
\hline
$ee$ & 225 & 210 & 205 \\
$e\mu$ & 210 & 195  &  190 \\
$\mu\mu$ & 245& 220  & 205\\
\hline\hline
\end{tabular}
\label{tab:hm}
\end{table}

In summary, we present a search for new physics in events with two high $p_T$
 leptons of the same electric charge using data with an integrated
 luminosity of 6.1 fb$^{-1}$. The observed data are consistent with
 standard model predictions. We set 95\% confidence level lower limits on the mass of
 doubly-charged scalars decaying to like-sign dileptons, $m_{H^{\pm\pm}}  > $
 190-245 GeV/$c^2$, depending on the decay mode and coupling.

We thank the Fermilab staff and the technical staffs of the
participating institutions for their vital contributions. This work
was supported by the U.S. Department of Energy and National Science
Foundation; the Italian Istituto Nazionale di Fisica Nucleare; the
Ministry of Education, Culture, Sports, Science and Technology of
Japan; the Natural Sciences and Engineering Research Council of
Canada; the National Science Council of the Republic of China; the
Swiss National Science Foundation; the A.P. Sloan Foundation; the
Bundesministerium f\"ur Bildung und Forschung, Germany; the Korean World Class University Program, the National Research Foundation of Korea; the Science and Technology Facilities Council and the Royal Society, UK; the Institut National de Physique Nucleaire et Physique des Particules/CNRS; the Russian Foundation for Basic Research; the Ministerio de Ciencia e Innovaci\'{o}n, and Programa Consolider-Ingenio 2010, Spain; the Slovak R\&D Agency; and the Academy of Finland. 

\begin{thebibliography}{99}
\bibitem{coordinates}
CDF uses a cylindrical coordinate system with
  the $z$ axis along the proton beam axis. Pseudorapidity is $\eta
  \equiv -\ln(\tan (\theta/2))$, where $\theta$ is the polar angle
  relative to the proton beam direction, and $\phi$ is the azimuthal angle  while $p_T = |p|\ \sin\theta$, $E_T = E\ \sin\theta$.
\bibitem{metdef} 
Missing transverse momentum, $\missET$, is defined as the magnitude of the vector $ -\sum_i E_T^i \vec n_i$ where $E_T^i$ are the magnitudes of transverse energy contained in each calorimeter tower $i$, and $\vec n_i$ is the unit vector from the interaction vertex to the tower in the transverse ($x,y$) plane.
\end{thebibliography}

\end{document}